\documentclass[aps,pra,twocolumn,superscriptaddress,amsmath,amssymb]{revtex4-1}
\pdfoutput=1
\usepackage{graphicx}
\usepackage{amsmath}
\usepackage{color}
\usepackage[normalem]{ulem}
\def\bm#1{\mathbf{#1}}

\begin{document}

\title{Dressing the Orbital Feshbach Resonance using single-manifold Raman scheme}

\author{Zhen Han}
\email{jhenhan@mail.ustc.edu.cn}
\affiliation{Department of Physics, University of Science and Technology of China, Chinese Academy of Sciences, Hefei, Anhui, 230026, China}
\author{Tian-Shu Deng}
\email{sky2010@mail.ustc.edu.cn}
\affiliation{Key Laboratory of Quantum Information, University of Science and Technology of China, Chinese Academy of Sciences, Hefei, Anhui, 230026, China}
\affiliation{Synergetic Innovation Center of Quantum Information and Quantum Physics, University of Science and Technology of China, Hefei, Anhui 230026, China}

\date{\today}
\begin{abstract}
The recently discovered Orbital Feshbach Resonance (OFR) offers the possibility of tuning the interaction between alkaline earth(-like) metal atoms with magnetic field. Here, we introduce a single-manifold Raman scheme to dress the OFR, which allows us to tune the interaction with the optical field and it is readily realizable in experiment. We demonstrate the scattering resonance could be shifted by the dressing Raman laser using few-body and many-body mean-field calculation, which give rise to an optical dependent two-body bound state and Raman coupling induced BCS-BEC crossover in the BCS-type mean field theory. Besides, we also discuss the application of single-manifold Raman scheme in Kondo research by writing down a Kondo lattice model.
\end{abstract}
\maketitle
\section{Introduction}

  Magnetic Feshbach Resonance(MFR) realize the tunability of the two-body interaction and it has always been a powerful tool in cold atoms system~\cite{MFR1,MFR2,MFR3,MFR4,MFR5,MFR6}. However, for alkaline-earth(-like) metal atoms, the outer shell is fully occupied and their total electron spin is zero. Thus, it has been considered difficult to construct two different potentials in alkali-earth(-like) metal atoms at ground state. Orbital Feshbach Resonance(OFR) was recently proposed to solve this problem and it significantly enriches the study of strongly interacting systems~\cite{OFR1INTERACTIONOFYB,OFR2,OFR3,OFR4,OFR5YBPARA4,OFR6YBPARA1}. In the OFR system, two alkali-earth(-like) metal atoms are prepared in two different electronic (orbital) states $^1S_0$ (stable) and  $^3P_0$ (long-lived metastable state), associated with two nuclear spins. It involves four hyperfine states $|g\uparrow(\downarrow)\rangle$ and $|e\uparrow(\downarrow)\rangle$ as shown in Fig.1(a). Here, $|g\uparrow;e\downarrow\rangle$ and $|g\downarrow;e\uparrow\rangle$ are defined as closed channel and open channel respectively. The two-body interaction at short ranges is diagonal in two anti-symmetrized basis $|\pm\rangle=(|ge\rangle\pm|eg\rangle)(\uparrow\downarrow\rangle\mp\downarrow\uparrow\rangle)$, which results from the fact that when the electronic states are anti-symmetric(symmetric), the nuclear spin states must be symmetric(anti-symmetric) due to Fermi atoms statistics~\cite{OFR1INTERACTIONOFYB}. Magnetic field could be used to control the Zeeman energy shift differential so that the system could be tuned to reach a scattering resonance.

\begin{figure}
  \centering
  \includegraphics[width=8cm]{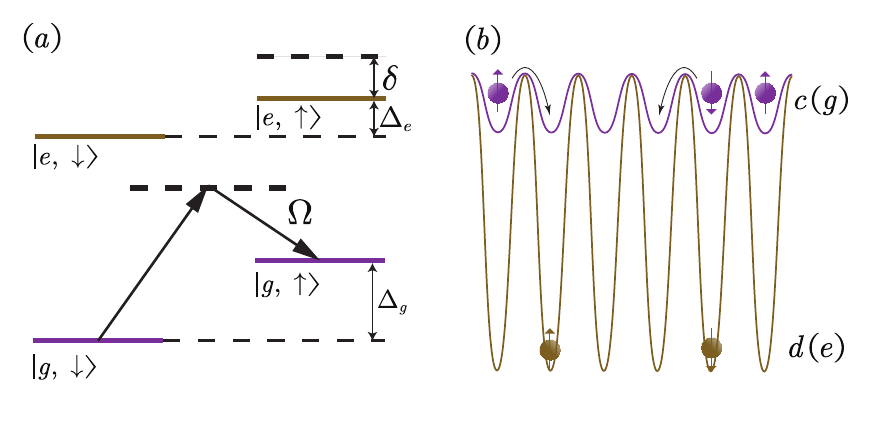}\\
  \caption{(a) Single-manifold scheme: imposing optical coupling to typical four energy levels involved with an OFR, g and e indicates $^1S_0$ and $^3P_0$ electronic states, and for simplicity, we take the two nuclear spin states $|\uparrow\rangle$ and $|\downarrow\rangle$ with magnetic angular momentum $m_F$ and $m_F+1$. (b) Kondo Lattice illustration: The green(purple) balls are in a c(d) state, which is a mixture of two substates of e(d) state. Atoms of c state are trapped in a deep lattice. They are localized, and have a lower density, while atoms of c state are trapped in a shallow lattice and are itinerant. The arrows denote the helicity branches.}
  \label{fig:fig0}
\end{figure}
In a previous study by one of the authors~\cite{PREVIOUSWORK}, we found that the optical field could dress the OFR and tune the scattering resonance location. We have given two different schemes, Raman scheme and Rabi scheme. In the Rabi scheme we directly drive the clock transition $^1S_0-^3P_0$ and couple different electronic states with the same nuclear spin. However, it induce an inevitable momentum transfer, and in this regime we cannot study Kondo effect due to the electronic states mixing. The Raman scheme couples different nuclear spin states with the same orbital index and the realization may involve two pairs of Raman lasers, which may suffer from strong heating in experiment~\cite{HEATING}. So we now propose a single-manifold Raman scheme as shown in Fig.1. Here, we apply one pair of Raman laser to couple two substates with different nuclear spin of ground state $^1S_0$ and we choose 1013nm Raman laser to assure the corresponding AC polarization of $^3P_0$ vanishes~\cite{AC}.
Also, it could avoid the momentum transfer by adjusting two beams of Raman laser parallel.

In this paper, we demonstrate the feasibility of the single-manifold Raman coupling in a dressed OFR system by studying both few-body  and many-body physics. For the few-body physics, we calculate the effective two-body scattering length and bound state energy. And we found that two branches of bound state energy corresponds to two different interaction potentials parameterized with $a_s^+$, $a_s^-$ respectively, and  we  could have a simple understanding for two-body bound state energy and wave function using the nuclear spin singlet and triplet basis. For the many body physics, we follows the BCS-type mean-field approach to study the BCS-BEC crossover near a dressed OFR. The order parameter and chemical potential as a function of Rabi frequency indicate that Raman coupling could tune the system through the crossover region and into the BCS regime. And then we also discuss the potential usage of our scheme to study the Kondo effect by writing down a Kondo lattice model with spin-exchange interaction between localized impurities and itinerant fermions.

The remainder of this paper is organized as follows. In Sec.II, we study the two body physics in the single-manifold Raman scheme, which includes a scattering length calculation in II.A and the derivation of two body bound state in II.B. In Sec.III we study the BEC-BCS crossover with the mean-field theory. In Sec. IV we give a kondo lattice model under this regime.

\section{FEW-BODY CALCULATION}

\subsection{OFR in alkaline-earth-like atoms}
 As illstrated in Fig.1(a), we apply two co-propagator laser to couple $|g\uparrow\rangle$ with $|g\downarrow\rangle$. The non-interacting Hamiltonian can be written as
\begin{align}\label{fb_imp_h_0}
&H_0\nonumber\\
&=\hat{T}\left| e\uparrow ;g\downarrow \right> \left< e\uparrow ;g\downarrow \right|+\left( \hat{T}+\delta \right) \left| e\downarrow ;g\uparrow \right> \left< e\downarrow ;g\uparrow \right|
\nonumber\\
&+\left( \hat{T}+\delta /2 \right) \left( \left| e\downarrow ;g\downarrow \right> \left< e\downarrow ;g\downarrow \right|+\left| e\uparrow ;g\uparrow \right> \left< e\uparrow ;g\uparrow \right| \right)
\nonumber\\
&+\Omega \left( \left| g\downarrow \right> \left< g\uparrow \right|+H.c. \right),
\end{align}
where $\left| e,\sigma _1;g,\sigma _2 \right> =\left| e,\sigma _1 \right>  \left| g,\sigma _2 \right> $ denote the two atoms state that the $e$ state atom is in the $\left|\sigma_{1} \right> $ nuclear spin state while $g$ state atom in the $\left| \sigma _2 \right> $ state ($\sigma _1,\sigma _2={ \uparrow ,\downarrow } $). $\hat{T}=\hbar^2\nabla ^2/m$ is the kinetic operator. $\Omega$ is the effective  Rabi-frequency for the single-manifold raman process. $\delta$ is the energy detune in different channels, given by $\delta =\Delta _g-\Delta _e$, with $\Delta _{g/e}$ the Zeeman shift in $g/e$ state manifold. We have $\Delta _{g/e}=g_{g/e}\mu _BB$, with $g_{g/e}$ the Lande factors, $\mu_B$ the Bohr magneton and $B$ the external magnetic field. Here we have made the approximation of no momentum transfer during the Raman process, which can be validiated in the case of co-propagating Raman lasers.

Since the appropriate basis for the description of the scattering is two anti-symmetrized basis $|\pm\rangle$, the Huang-Yang
pseudo-potential has the form of $H_{\text{int}}=\sum_{j=\pm}a_s^j\delta \left( \boldsymbol{r} \right) \frac{\partial}{\partial r}\left( r\cdot \right)|j\rangle\langle j|$~\cite{HUANGYANG}. Then we transform it into nuclear spin and electronic states basis, it derives
\begin{align}\label{fb_imp_h_int}
H_{\text{int}}&=U_{1}(\left| e\uparrow ;g\downarrow \right> \left< e\uparrow ;g\downarrow \right|+\left| e\downarrow ;g\uparrow \right> \left< e\downarrow ;g\uparrow \right|)
\nonumber\\
&+U_{2}(\left| e\uparrow ;g\downarrow \right> \left< e\downarrow ;g\uparrow \right|+\left| e\downarrow ;g\uparrow \right> \left< e\uparrow ;g\downarrow \right|).
\end{align}

The coefficient $U_{1}$ and $U_{2}$ are the contact interaction potential relating to scattering length between channels: $U_{i}=a_i\delta \left( \boldsymbol{r} \right) \frac{\partial}{\partial r}\left( r\cdot \right) $. And $a_{1,2}$ is connected with $a_s^{\pm}$ as $a_{1}=\left( a^{+}_{s}+a^{-}_{s} \right) /2$ and $a_{2}=\left( a^{+}_{s}-a^{-}_{s} \right) /2$ \cite{PREVIOUSWORK}.

The non-interacting Hamiltonian with inter-channel coupling can be diagonalized under the basis $\{\left| g\alpha \right> \left| e\beta \right> \}$ $(\alpha,\beta=1,2)$, and the Bogoliubov transformation gives $\left| g1 \right> =\cos \theta \left| g\uparrow \right> -\sin \theta \left| g\downarrow \right> $, $\left| g2 \right> =\sin \theta \left| g\uparrow \right> +\cos \theta \left| g\downarrow \right> $, $\left| e1 \right> =\left| e\uparrow \right> $ and $\left| e2 \right> =\left| e\downarrow \right> $, with $\tan \theta =( \delta /4+\sqrt{( \delta /4 ) ^2+\Omega ^2} ) /\Omega $.  The diagonalized Hamiltonian is written as $H_0=\Sigma _{\alpha ,\beta}(-\hbar^2\nabla ^2/m+ \epsilon _{g,\alpha}+\epsilon _{e,\beta} ) \left| \alpha \beta \right> \left< \alpha \beta \right|$. Here, $\left| \alpha \beta \right> $ is the shorthand notation of $\left| g\alpha \right> \left| e\beta \right> $ , and the single-particle dispersion $\epsilon _{g,1}=\delta /4-\sqrt{( \delta /4 ) ^2+\Omega ^2}$, $\epsilon _{g,2}=\delta/4+\sqrt{( \delta/4 ) ^2+\Omega ^2}$, $ \epsilon _{e,1}=0$, $\epsilon _{e,2}=\delta /2$. The two-body energy threshold is given by $E_{\text{th}}=\epsilon _{g,1}+\epsilon _{e,1}=\delta /4-\sqrt{( \delta /4 ) ^2+\Omega ^2}$.

\begin{figure}  [tbp]
\centering
\includegraphics[width=8cm]{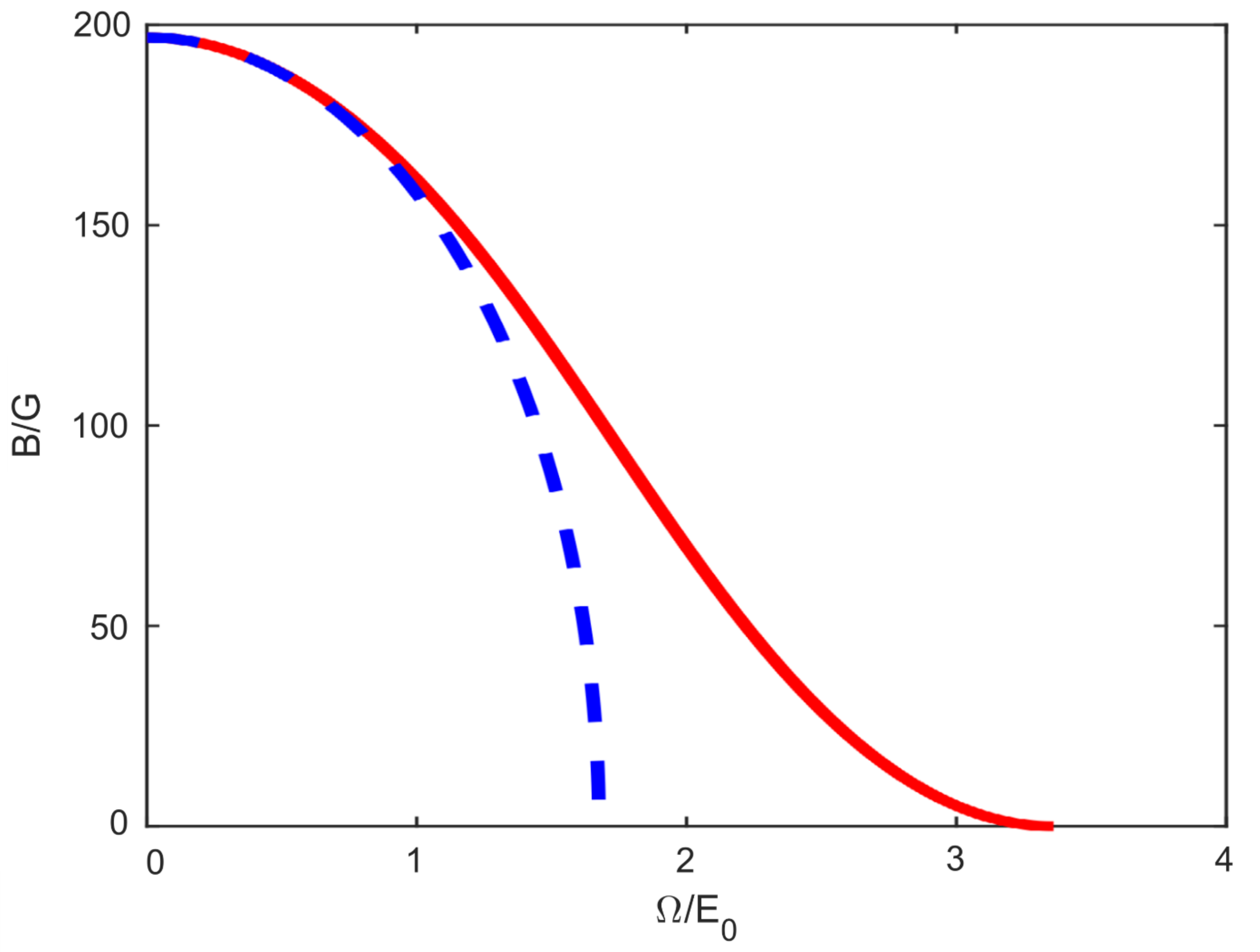}
\caption{The two body resonance point on the $\Omega-B$ plane where scattering length of the lowest-energy channel diverges in single- manifold Raman scheme (red solid) and Raman scheme (blue dash). The result of Raman scheme is computed according to the expression of scattering length in[] . Here the unit of energy is chosen as $E_0=\hbar^2k_{0}^{2}/2m$ and $k_{0}$ is the wave vector of the 578nm clock transition $^1S_0\rightarrow ^3P_0$ in $^{173}\text{Yb}$. We have taken the parameters of $^{173}\text{Yb}$ for our calculations, with $a_{s}^{-}=219.5a_0$, $a_{s}^{+}=1900a_0$, where $a_{0}$ is the Bohr's radius,$g_g\mu _B=2\pi {\hbar }\times \text{207.15Hz/G}$, $g_e\mu _B=2\pi {\hbar }\times \text{93.78Hz/G} $~\cite{OFR6YBPARA1,OFR5YBPARA4,YBPARA2,YBPARA3}. For single-manifold Raman scheme, in the absence of magnetic field, two body resonance occur at $\Omega =3.36E_0.$}
\label{fig:fig1}
\end{figure}

 In order to study the scattering resonance problem in this model, we need to practice the partial-wave analysis and calculate lowest energy scattering amplitude. Using the usual partial-wave expansion, we can write wave function as $\left| \psi _{\mathbf{k}}\left( \boldsymbol{r} \right) \right> =\left[ e^{i\mathbf{k}\cdot \mathbf{r}}+f_{11}\left( \mathbf{k} \right) \frac{e^{ikr}}{r} \right] \left|  11 \right> +\sum_{\alpha ,\beta \ne \left( \text{1,}1 \right)}{f_{\alpha \beta}\left( \mathbf{k} \right) \frac{e^{-\kappa _{\alpha \beta}r}}{r}\left| \alpha \beta \right>}$, with $\kappa _{\alpha \beta}=\sqrt{m\Delta _{\alpha \beta}/{\hbar }^2-k^2}$, $\Delta _{\alpha \beta}=\epsilon _{g,\alpha}+\epsilon _{e,\beta}-\epsilon _{g,1}-\epsilon _{e,1}$, and $f_{\alpha\beta}$ the scattering amplitude. Substituting it into the Schr\"{o}dinger's equation and we could get the solution of scattering amplitude $f_{ij}$. The scattering length for lowest incident channel could be taken from the limit $k\rightarrow 0$, and it yields that
\begin{align}\label{as}
a_{s}^{( 11 )}&=-\lim_{k\rightarrow 0}f_{11}( \mathbf{k} )
\nonumber\\&
=\frac{2a_{oo}\sin ^2\theta +\sqrt{2}a_1R\sin ^2\theta}{2-a_2\cos ^2\theta -a_1\sqrt{R_1R_2}\cos ^4\theta -a_3\sin ^2\theta},
\end{align}
where $R_1=m\sqrt{\delta ^2/16+\Omega ^2}/{\hbar }^2$, $R_2=m\delta /{\hbar }^2$, $R=\sqrt{R_1}\sin ^2\theta +\sqrt{R_2}\cos ^2\theta $, $a_1=a_{co}^2-a_{cc}a_{oo}$, $a_2=\sqrt{2}a_{oo}\sqrt{R_1}+\sqrt{2}a_{cc}\sqrt{R_2}$, $a_3=\sqrt{2}a_{cc}\sqrt{R_1}+a_1R_1\cos ^2\theta $.
We can see $a_{s}^{( 11 )}$ and its denominator is dependent on $\delta$ and $\Omega$, and since the two-body resonance occur when the scattering length diverges, both the resonance point and scattering length depends on these two parameters.

Since $\delta$ is decided by the magnetic field, we can locate the resonance points on the $\Omega-B$ plane as we did in Fig.\ref{fig:fig1} using $^{173}\text{Yb}$ for example. Note that when the dressing parameter is zero, the resonance points of ordinary Raman scheme and single-manifold Raman scheme coincide, since two models are the same when $\Omega=0$. We can see that the effective Rabi-frequency required in the single-manifold Raman scheme to reach two-body resonance is larger than the one in Raman scheme by a factor up to 2. It suggest the feasiblity of implementing single-manifold Raman scheme with the similiar experiment condition in Raman scheme, while one less Raman process in single-manifold version permits simplification of experiment setup concerning Raman lasers.

\subsection{Bound state energy and wavefunction}

In single-manifold Raman scheme, when the magnetic field is zero and thus there is no Zeeman shift in clock-state manifold, the two-body bound state energy and the channel amplitude have a quite simple relation with scattering length in different channels and the dressing parameter. For better understanding of the two-body bound state in this model, we start from the second quantized form of Hamiltonian $H=H_0+H_{\text{int}}$ in the absence of magnetic field. {The non-interacting term has the form of
\begin{align}\label{H_int_0_2q}
H_0&=\sum_{j,\sigma ,\mathbf{k}}{\epsilon _{\mathbf{k}}a_{j,\downarrow ,\mathbf{k}}^{\dag}a_{j,\downarrow ,\mathbf{k}}}+\Omega \sum_{\mathbf{k}}{\left( a_{g,\uparrow ,\mathbf{k}}^{\dag}a_{g,\downarrow ,\mathbf{k}}+H.c. \right)}
\end{align}
where $a_{j,\sigma ,\boldsymbol{k}}^{\dag}( a_{j,\sigma ,\boldsymbol{k}} ) $ creates (annihilates) an atom in the state $\left| j,\sigma \right> $
($j=\{g,e\},\sigma=\{\uparrow,\downarrow\}$) with momentum $\hbar \boldsymbol{k}$.
$\epsilon_{\boldsymbol{k}}=\hbar^2k^2/2m$, and again $\Omega$ is the dressing parameter.
 The interaction term could be written as 
 \begin{equation}\label{H_int_0_2q1}
   H_{{\text{int}}}=\sum_{\mathbf{q},\bm{k},\bm{k}',\sigma=\pm}g_{\sigma}A_{\sigma}^{\dagger}(\mathbf{q},\bm{k})A_{\sigma}(\mathbf{q},\bm{k}').
\end{equation}
The coupling constant $g_{\pm}$ is associated with the scattering length $a_{\pm}$ as: $\text{1/}g_{\pm}=\text{1/}g_{\pm}^{p}-\Sigma _{\mathbf{k}}\text{1/}2\epsilon _{\mathbf{k}}$ and $g_{\pm}^{p}=4\pi {\hbar }^2a^{\pm}_{s}/m$.
And, we define
\begin{align}
A^{\dag}_+\left(\mathbf{Q}, \mathbf{k} \right) &=\frac{1}{\sqrt{2}}\left( a_{e,\uparrow ,\mathbf{Q}-\mathbf{k}}^{\dag}a_{g,\downarrow ,\mathbf{k}}^{\dag}-a_{e,\downarrow ,\mathbf{Q}-\mathbf{k}}^{\dag}a_{g,\uparrow ,\mathbf{k}}^{\dag} \right),
\\A^{\dag}_-\left( \mathbf{Q}, \mathbf{k} \right) &=\frac{1}{\sqrt{2}}\left( a_{e,\downarrow ,\mathbf{Q}-\mathbf{k}}^{\dag}a_{g,\uparrow ,\mathbf{k}}^{\dag}+a_{e,\uparrow ,\mathbf{Q}-\mathbf{k}}^{\dag}a_{g,\downarrow ,\mathbf{k}}^{\dag} \right) ,
\\A^{\dag}_1\left( \mathbf{Q}, \mathbf{k} \right) &=a_{e,\uparrow ,\mathbf{Q}-\mathbf{k}}^{\dag}a_{g,\uparrow ,\mathbf{k}}^{\dag},
\\A^{\dag}_{-1}\left( \mathbf{Q}, \mathbf{k} \right) &=a_{e,\downarrow ,\mathbf{Q}-\mathbf{k}}^{\dag}a_{g,\downarrow ,\mathbf{k}}^{\dag}.
\end{align}
Here,$A^{\dag}_+\left( \mathbf{Q}, \mathbf{k} \right)$ is creation operator of one singlet state with total momentum $\mathbf{Q}$ and $g$ state atom momentum $\mathbf{k}$, and $A^{\dag}_-\left( \mathbf{Q}, \mathbf{k} \right)$, $A^{\dag}_{\pm{1}}\left( \mathbf{Q}, \mathbf{k} \right)$ are  three triplet states. These form the basis of short-range interaction. Then we can write down the two-body bound state wave function as
\begin{align}
\left| \Psi \right> _{\mathbf{Q}}=\sum_{\mathbf{k}}{\left( \psi _+\left( \mathbf{k} \right) A_{+}^{\dag}\left( \mathbf{Q}, \mathbf{k} \right) +\psi _-\left( \mathbf{k} \right) A_{-}^{\dag}\left( \mathbf{Q}, \mathbf{k}\right) \right.}
\nonumber\\
\left. +\psi _1\left( \mathbf{k} \right) A_{1}^{\dag}\left( \mathbf{Q}, \mathbf{k} \right) +\psi _{-1}\left( \mathbf{k} \right) A_{-1}^{\dag}\left( \mathbf{Q}, \mathbf{k} \right) \right) \left| vac \right>
\end{align}
with the bound state wavefunction $\psi _{+,-,\text{1,}-1}$}

To calculate the bound state energy, we derive the equation for wavefunctions by comparing the coefficient of $A_{\lambda}^{\dag}$ ($\lambda=+,-,1,-1$) in the Schr\"{o}dinger's equation $\left( H_0+H_{\text{int}} \right) \left| \Psi \right> _{\mathbf{Q}}=E\left| \Psi \right> _{\mathbf{Q}}$:
\begin{align}\label{2bsc}
\left( \begin{matrix}
	T-E&		0&		\frac{\Omega}{\sqrt{2}}&		-\frac{\Omega}{\sqrt{2}}\\
	0&		T-E&		\frac{\Omega}{\sqrt{2}}&		\frac{\Omega}{\sqrt{2}}\\
	\frac{\Omega}{\sqrt{2}}&		\frac{\Omega}{\sqrt{2}}&		T-E&		0\\
	-\frac{\Omega}{\sqrt{2}}&		\frac{\Omega}{\sqrt{2}}&		0&		T-E\\
\end{matrix} \right) \left( \begin{array}{c}
	\psi _+\left( \mathbf{k} \right)\\
	\psi _-\left( \mathbf{k} \right)\\
	\psi _1\left( \mathbf{k} \right)\\
	\psi _{-1}\left( \mathbf{k} \right)\\
\end{array} \right) =\left( \begin{array}{c}
	F_{+}\\
	F_{-}\\
	0\\
	0\\
\end{array} \right)
\end{align}
with $F_{\pm}=-g_{\pm}\Sigma _{\mathbf{k}}\psi _{\pm}\left( \mathbf{k} \right)$ and $T=\epsilon _{\mathbf{k}}+\epsilon _{\mathbf{Q}-\mathbf{k}}$. Then we can get closed equations for $\psi _{\pm}$:
\begin{align}
\psi _{\pm}\left( \mathbf{k} \right) =-\frac{F_{\pm}}{\frac{\Omega ^2}{T-E}-\left(T-E \right)}
\end{align}
In absence of the magnetic field, the energy difference between spin states in $g$ state is zero and thus the single-manifold Raman scheme should have no momentum transfer if the lasers are counter-propagatingly configured. For simplicity, we assume $\mathbf{Q}=0$. Thus we can solve the closed equation by summing (\ref{2bsc}) over $\mathbf{k}$ and eventually get two branch of possible energy solutions:
\begin{align}
E_{\pm}=-\frac{\hbar^2}{ma_{s}^{\pm2}}-\frac{ma_{s}^{\pm2}}{4\hbar^2}\Omega ^2.
\end{align}
It's a quite simple result since each branch depend on $a^{+}_{s}$ or $a^{-}_{s}$, respectively.

The bound state energy reaches the two-body energy threshold $E_{\text{th}}=-\Omega$ when $\Omega=\frac{2{\hbar }^2}{ma_{s}^{\pm2}}$. The deeper branch, related to - channel scattering length $a_s^-$, is far detuned from the threshold. Thearfore, in the following discussion, we mainly concern about the shallow branch. In the case of $^{173}\text{Yb}$, $E_+=E_{\text{th}}=-\Omega$ gives $\Omega=\frac{2{\hbar }^2}{ma_{s}^{+2}}=3.36E_{0}$, which is consist with the position of resonance point of single-manifold Raman scheme when $B=0$ in Fig. \ref{fig:fig1}.

By substituting the solutions of bound state energy into the Schr\"{o}dinger's equation Eq.(\ref{2bsc}), we can derive the wave function as (we consider the $E_+$ branch here):
\begin{align}
\psi _+\left( \mathbf{k} \right) &=\frac{C}{E_++\Omega -2\epsilon _{\mathbf{k}}}+\frac{C}{E_+-\Omega -2\epsilon _{\mathbf{k}}},
\\
\psi _-\left( \mathbf{k} \right) &=0,
\\
\psi _1\left( \mathbf{k} \right) &=\frac{\sqrt{2}}{2}\frac{C}{E_+-\Omega -2\epsilon _{\mathbf{k}}}-\frac{\sqrt{2}}{2}\frac{C}{E_++\Omega -2\epsilon _{\mathbf{k}}},
\\
\psi _{-1}\left( \mathbf{k} \right) &=\frac{\sqrt{2}}{2}\frac{C}{E_++\Omega -2\epsilon _{\mathbf{k}}}-\frac{\sqrt{2}}{2}\frac{C}{E_+-\Omega -2\epsilon _{\mathbf{k}}},
\end{align}
where $C$ is the normalization constant and we could determine its value from $\sum_{\lambda=\pm,\pm1}\int{\left|\psi _{\lambda}\left( \mathbf{k} \right)  \right|^2d^3\mathbf{k}}=1$. Diagonalizing the non-interacting Hamiltonian gives the two quasi-particles eigenenergy $2\epsilon _{\mathbf{k}}\pm \Omega$, and here we could see the wavefunction $\psi _{\lambda}$ diverges when the molecule energy coincides with this threshold.

The population can be given by $W^{\lambda}=\int{\left|\psi _{\lambda}\left( \mathbf{k} \right)  \right|^2d^3\mathbf{k}}$ to indicate the possiblity for two atoms be in the one singlet and three triplet states. We find it surprisingly depend quadratically on $\Omega$:
\begin{align}\label{wfw}
&W_+=1-2\left( \frac{m{a_s^+}^2}{4\hbar^2} \right) ^2\Omega ^2\nonumber\\&W_-=0\nonumber\\
&W_1=W_{-1}=\left( \frac{m{a_s^+}^2}{4\hbar^2} \right) ^2\Omega ^2
\end{align}
Notice that $\left<vac\right|A_{-}\left(\mathbf{k}_{1}\right)(H_{0}+H_{\text{int}})A_{+}^{\dag}\left(\mathbf{k}_{2}\right)\left|vac\right>=0$ which means either optical coupling or interaction term cannot mix the $\pm$ channels. Thus, for the shallow branch $E_+$, channel $-$ wave function vanishes.
%Notice that the transition matrix element between $+$ and $-$ channel provided by the inter-channel dressing is
%$\left< vac \right|A_-\left( \mathbf{k}_1 \right)  \Omega\left( \sum_{\mathbf{k}}{a_{g,\uparrow ,\mathbf{k}}^{\dag}a_{g,\downarrow ,\mathbf{k}}} +H.c. \right) A_{+}^{\dag}\left( \mathbf{k}_2 \right) \left| vac \right> =0$. Also, the interaction Hamiltionian is diagonalized under the singlet-triplet basis. These two fact explained the absence of atoms in $-$ channel. 
When $\Omega=0$, all atoms are in the $+$ channel. As the inter-channel dressing is turned on, $\pm1$ channels are equally occupied due to the symmetry. When $\Omega$ reaches resonance point, the populations of the triplet and singlet states equal, and thus the interaction is most intense.

\begin{figure}  [tbp]
\centering
\includegraphics[width=8cm]{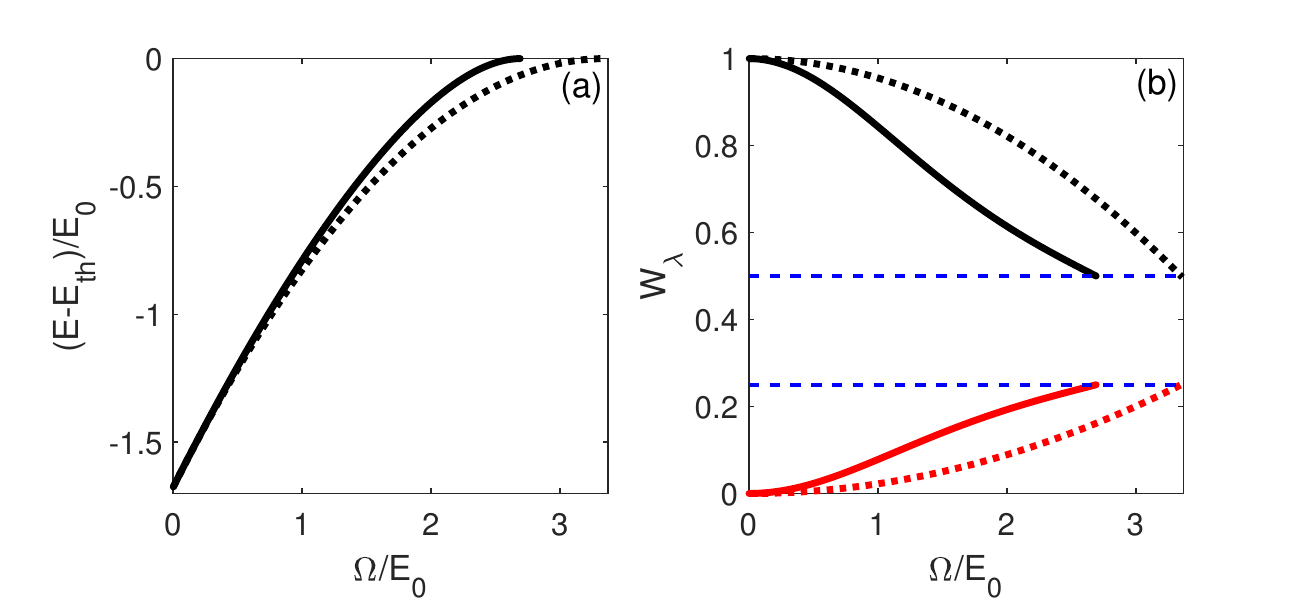}
\caption{(a) Two body bound state energy relatives to the bound state energy threshold. (b) Populations in different channels. Here in both (a) and (b) solid line stands for four interacting channels model and dot line stands for two interacting channels model. In (b), black line is the population in + channel and red line is the one in 1 or -1 channel. We choose the most shallow branch of solution of $^{173}\text{Yb}$ as example here. For the bound state energy, the four interacting channels model reaches the resonance point at $\Omega=2.69E_{0}$. For the population,in both models the population in - channel is always 0, and the ones in 1 and -1 channel always equal. In the four interacting channels model the population no longer have quadratic relation with the coupling parameter, but population in + channel still reaches half at the resonance point.}
\label{fig:figchan}
\end{figure}

We considered the short-range interaction in two channels, + and - channels, in most part of our calculation in this work and previous work of one of the authors \cite{PREVIOUSWORK}. In fact, a more complete consideration of interaction should include all four channels , having the form of $g_+A_{+}^{\dag}A_++\Sigma _{\sigma =-,\text{1,}-1}g_-A_{\sigma}^{\dag}A_{\sigma}$ \cite{KONDOHUIZHAI2CIR}. However, the reduction of interacting channels as approximation doesn't change the result much in our study. For concreteness, here we give the two-body bound state energy and the population of $^{173}\text{Yb}$ in the four interacting channels model in Fig.\ref{fig:figchan}, and it turns out to be only slightly different with the ones in the two interacting channels model we used above.

\section{BCS-BEC CROSSOVER}
To probe on the multi-body properties of the single-manifold Raman scheme, we calculated the BCS-BEC crossover using the BCS-type mean field approach in the $B=0$ case. In detail,  we defined two pairing parameters,  $\Delta_{\pm}=\left(g_{\pm}/2\right)\left<\sum_{\bm{k}}A_{\pm}(0,\bm{k})\right>$. Then we expanded $H_{\text{int}}$ to the first order of fluctuation $\delta_{\pm}=\sum_{\bm{k}}\left(A_{\pm}(0,\bm{k})-\left<A_{\pm}(0,\bm{k})\right>\right)$ to get the mean-field interacting Hamiltonian:
\begin{align}
H_{\text{int}}&=\frac{g_+}{2}\left( \delta _{+}^{\dag}+\frac{2}{g_{+}}\Delta_{+} \right) \cdot H.c.+\frac{g_-}{2}\left( \delta _{-}^{\dag}+\frac{2}{g_{-}}\Delta_{-} \right) \cdot H.c.
\nonumber\\
&\approx\sum_{\bm{k}}\left(\Delta_{+}^{\dag}A_{+}(0,\bm{k})+\Delta_{-}^{\dag}A_{+}(0,\bm{k})+H.c.\right)\nonumber\\
&-\frac{2}{g_+}\Delta _+^2-\frac{2}{g_-}\Delta _-^2
\end{align}
Due to the omittable momentum transfer in single-manifold Raman scheme, here we adopted zero total momentum case. Thus the thermodynamic potential $\mathcal{K}=\left< H-\mu N \right>$ can be written as
\begin{align}
\mathcal{K}&=\nonumber\\
&\sum_{\mathbf{k}}{\Psi _{\mathbf{k}}^{\dag}}\left( \begin{matrix}
	\epsilon _{\mathbf{k}}-\mu&		0&		0&		\Delta _--\Delta _+\\
	0&		\epsilon _{\mathbf{k}}-\mu&		\Delta _++\Delta _-&		0\\
	0&		\Delta _++\Delta _-&		-\epsilon _{\mathbf{k}}+\mu&		-\Omega\\
	\Delta _--\Delta _+&		0&		-\Omega&		-\epsilon _{\mathbf{k}}+\mu\\
\end{matrix} \right) \Psi _{\mathbf{k}}
\nonumber\\
&+\left( -\frac{2}{g_+}\Delta _+^2-\frac{2}{g_-}\Delta _-^2 \right)
\end{align}
with $\Psi _{\mathbf{k}}^{\dag}=\left( \begin{matrix}
	a^{\dag}_{e,\uparrow ,\mathbf{k}}&		a^{\dag}_{e,\downarrow ,\mathbf{k}}&		a_{g,\uparrow ,-\mathbf{k}}&		a_{g,\downarrow ,-\mathbf{k}}\\
\end{matrix} \right) $. From the BCS-type mean field approach, order parameters $\Delta _{\pm}$ and chemical potential $\mu$ can be extracted by directly solving thetwo gap equations $\partial \mathcal{K}/\partial \Delta _{\pm}=0$ and a number equation $\partial \mathcal{K}/\partial \mu =-N$.

\begin{figure} [tbp]
\centering
\includegraphics[width=8cm]{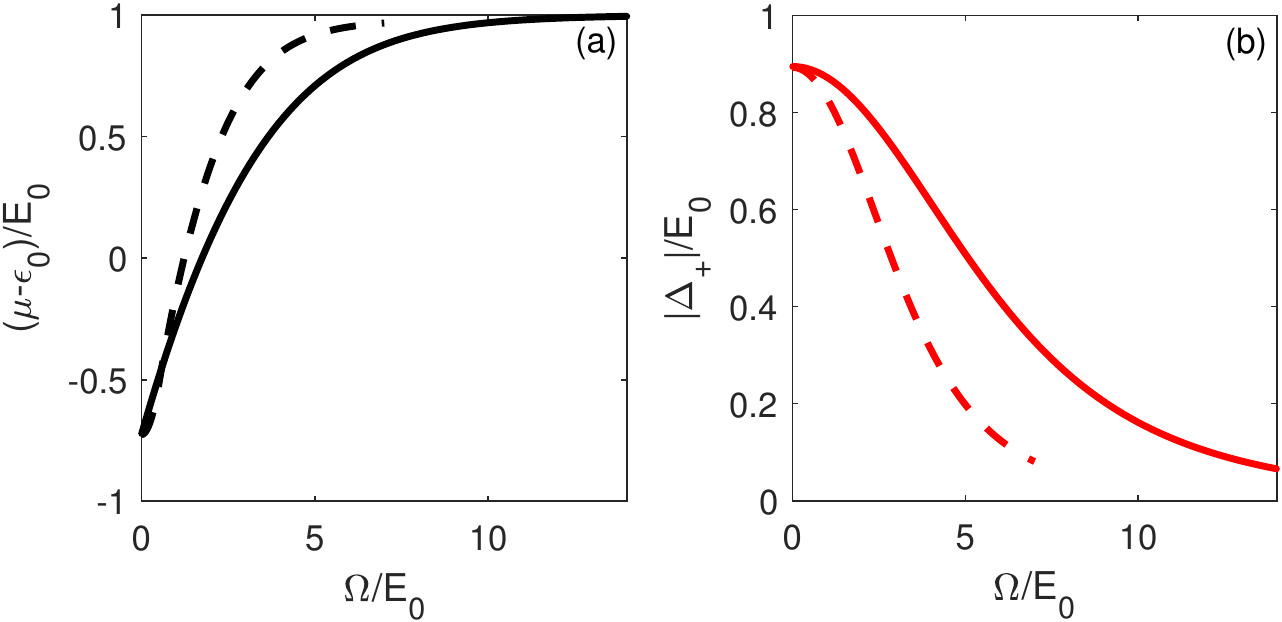}
\caption{(a) The chemical potential measured relative to the single-particle dispersion minimum $\epsilon_{0}=-\Omega$. (b) The pairing parameter $\Delta _+$. Here we fixed the phase of $\Delta_{+}$ as a real. It turns out that the other parameter $\Delta_-$ is always zero. We chose the density $n$ of $^{173}\text{Yb}$ as $n=k^{3}_{0}/3\pi^2$ with $k_0$ defined in the caption of Fig.\ref{fig:fig1}.}
\label{fig:bec}
\end{figure}

The results are shown in Fig.\ref{fig:bec}. As a comparison, we also plot the order parameter and chemical potential results of the two manifolds Raman scheme with the dashed line, and the two scheme matches with each other when $\Omega=0$.

We can clearly see the crossover in Fig.\ref{fig:bec}: as the $\Omega$ increasing, the pairing parameter $\Delta_{+}$ decreases and approaches 0, and the relative chemical potential change from negative to positive and approaches $E_0$. All the evidence shows that the system is approaching the BCS regime, and will eventually reach the deep BCS regime in the large-$\Omega$ limit. The crossover suggest that the single-manifold Raman scheme could be used to manipulate the interaction and thus the many-body properties in system of alkaline-earth like atoms.

\section{KONDO LATTICE MODEL}
Recently several researches focused on realizing Kondo physics in cold atom systems~\cite{KONDOHUIZHAI1,KONDOHUIZHAI2CIR,KONDOHUIZHAI3,KONDOHUIZHAI4,KONDOHUIZHAI5}. To demostrate the potential usage of single-manifold Raman scheme in OFR, we consider the lattice model of alkaline-like atoms. Suppose that the optical lattice is turned on and localizing  $\left| e \right> $ state atoms as impurity while leaving $\left| g \right> $ state atoms relatively itinerant as the Fermi sea. This can be archieved due to the difference in AC polarizability in two states. Then, considering the one dimensional tight-binging model under the nearest neighbour approximation, the non-interacting and interacting Hamiltonian is given by
\begin{align}\label{Kondo_H}
H_0&=\sum_{k,\sigma}{-2t\cos \left( kl \right) a_{g,\sigma ,k}^{\dag}a_{g,\sigma ,k}}\nonumber\\&+\Omega \sum_k{\left( a_{g,\uparrow ,k}^{\dag}a_{g,\downarrow ,k}+H.c. \right)},
\\
H_{\text{int}}&=\frac{J_0g_+}{2}\sum_{k,q}{A_{L+}^{\dag}\left( k \right) A_{L+}\left( q \right)}+\frac{J_0g_-}{2}\sum_{k,q}{A_{L-}^{\dag}\left( k \right) A_{L-}\left( q \right)}.
\end{align}
Here, we have defined $A_{L\pm}\left( k \right) =a_{g,\uparrow ,k}a_{e,\uparrow}\mp a_{g,\downarrow ,k}a_{e,\uparrow}$. $l$ is the lattice constant. $J_0$ is a constant that decided by the overlap between Wannier function of neighboor sites.To make the Hamiltonian consist with the ordinary form of Kondo Hamiltonian, we selected the rotated basis $\{c_{\uparrow ,k},c_{\downarrow ,k},d_{\uparrow ,k},d_{\downarrow ,k}\}$ with $\left[ \begin{array}{c}
	\begin{array}{c}
	c_{\uparrow ,k}^{\dag}\\
	c_{\downarrow ,k}^{\dag}\\
\end{array}\\
\end{array}\begin{array}{c}
	\begin{array}{c}
	d_{\uparrow}^{\dag}\\
	d_{\downarrow}^{\dag}\\
\end{array}\\
\end{array} \right] =\left[ \begin{matrix}
	\text{1/}\sqrt{2}&		\text{1/}\sqrt{2}\\
	-\text{1/}\sqrt{2}&		\text{1/}\sqrt{2}\\
\end{matrix} \right] \left[ \begin{array}{c}
	a_{g,\uparrow ,k}^{\dag}\\
	a_{g,\downarrow ,k}^{\dag}\\
\end{array}\begin{array}{c}
	a_{e,\uparrow}^{\dag}\\
	a_{e,\downarrow}^{\dag}\\
\end{array} \right] $.
The non interacting Hamiltonian now becomes $H_0=\Sigma _{k,\sigma}\epsilon _{c,k,\sigma}c_{\sigma ,k}^{\dag}c_{\sigma ,k}$. $\epsilon _{c,k,\sigma}$ is the single-particle dispersion of itinerant atoms. The interacting Hamiltonian can be written as
\begin{align}
H_{\text{int}}=\sum_k{J_+S^+c_{\downarrow ,k}^{\dag}c_{\uparrow ,k}+J_-S^-c_{\uparrow ,k}^{\dag}c_{\downarrow ,k}+\frac{1}{4}Un_{k}-J_zS^zs_z}
\end{align}
where we have defined $S^+=d_{\uparrow}^{\dag}d_{\downarrow}$, $S^-=d_{\downarrow}^{\dag}d_{\uparrow}$, $S^z=\text{1/}2\left( d_{\uparrow}^{\dag}d_{\uparrow}-d_{\downarrow}^{\dag}d_{\downarrow} \right)
$, $n_{k}=c_{\uparrow ,k}^{\dag}c_{\uparrow ,k}+c_{\downarrow ,k}^{\dag}c_{\downarrow ,k}$ and $s_z=\text{1/}2\left( c_{\uparrow ,k}^{\dag}c_{\uparrow ,k}-c_{\downarrow ,k}^{\dag}c_{\downarrow ,k} \right)
$.
The Kondo coefficients $J$ and $U$ are related to the coupling parameters $g_{\pm}$ through
\begin{align}
J_{\pm}=(\frac{g_-}{2}-\frac{g_+}{2})J_0,U=J_z=(g_++g_-)J_0
\end{align}

\section{CONCLUSION}
We showed that Orbital Feshbach Resonance can be realized with single-manifold Raman scheme just like the ordinary scheme while reducing the coupling lasers required. The simple quadratic relation between the two-body bound state energy and the inter-channel coupling parameter in this scheme is explored. Feasiblity to tuned from strongly to weakly interacting regime by this scheme is verified by the crossover. We give the corresponding Kondo-Lattice model for our system.

\section*{Acknowledgments}
We thank Wei Yi and Ren Zhang for helpful comments and discussions.

\end{document}